\documentstyle{disk}
\begin{document}

\title{%
Novae Crossing the Thermal Stability Line}

\author{
Alon RETTER,  Tim NAYLOR \\
{\it Dept. of Physics, Keele University, Keele,
Staffs., ST5 5BG, U.K., ar@astro.keele.ac.uk, timn@astro.keele.ac.uk} \\
Elia M. LEIBOWITZ \\
{\it School of Physics and Astronomy and the Wise Observatory, \\
Raymond and Beverly Sackler Faculty of Exact Sciences,\\
Tel-Aviv University, Tel Aviv, 69978, Israel, elia@wise.tau.ac.il
}
}

\maketitle

\section*{Abstract}

A method, based on the disc instability model, for testing the thermal 
stability of Cataclysmic Variables (CVs), is presented. It is shown that the 
border line between thermal stability and instability is crossed during some 
nova outbursts and decays, however it is not clear whether this is the general
behaviour. We suggest two new evolutionary scenarios for short orbital period 
CVs. One of them is the analogy for the 'modified hibernation scenario' and 
the other is an extension of the ideas of Mukai \& Naylor (1993) for long 
orbital period CVs. We conclude that the observations have not favoured one 
of the two models. Finally, we speculate the existence of a new type of 
nova - an AM CVn like nova.

\section{Introduction}

Osaki (1996) schematically divided the ($P_{orb}, \dot{M}$) plane into four 
regions. Long orbital period CVs (nova-likes and U Gem systems) are tidally 
stable, while short period systems (SU UMa and permanent superhump 
systems) are tidally unstable. CVs above his critical line (nova-likes and 
permanent superhumpers) are thermally stable, while the accretion discs of 
systems below this line (U Gem and SU UMa systems) are thermally unstable. 
Are these limits crossed?
It is clear that the tidal stability limit is hard to cross. In order to do 
so, the binary period has to be changed. However, a change in the mass 
transfer rate through the disc ($\dot{M}$) is much easier. Only an outburst 
is required for increasing $\dot{M}$. Indeed SU UMa systems are thermally 
unstable in quiescence, but the appearance of superhumps in their light 
curves seems to be evidence that they become quasi stable during 
superoutbursts. In this work, we concentrate on a different way of crossing 
this line - by outbursts (and decays) of non-magnetic classical novae.


\section{Models for CVs Evolution}

The spectrum of a typical old nova shows a strong continuum and emission 
lines. Nova-likes have similar features. Novae, nova-likes and dwarf novae 
share a similar binary configuration, namely a white dwarf and a red dwarf. 
It was thus suggested that these subclasses of CVs are evolutionarily 
connected. The 'hibernation scenario' (Shara 1989 for review) suggests that 
dwarf novae $\rightarrow$ nova-likes $\rightarrow$ novae $\rightarrow$ 
nova-likes $\rightarrow$ dwarf novae $\rightarrow$ hibernation $\rightarrow$ 
dwarf novae etc. However, it is now believed that the hibernation stage 
($\dot{M}$=0) might not exist at all (Livio 1989), thus
dwarf novae $\rightarrow$ nova-likes $\rightarrow$ novae $\rightarrow$ 
nova-likes $\rightarrow$ dwarf novae... The typical time scales for
the transitions were estimated as a few centuries. 

There is some observational evidence that supports this model. Livio 
(1989) listed a few old novae, which probably experienced dwarf nova outbursts
a few decades before the nova event. However, Robinson (1975) compared the 
magnitudes of 18 old novae with the values of their progenitors. He found no 
significant difference between these numbers, and concluded that all novae 
return to their pre-outburst luminosities. Although there are many light 
sources in old novae, it seems that their accretion discs are above the 
line (see section 4 and Warner 1995a). Since old novae are usually 
indistinguishable from nova-likes, it means that the outburst does not alter 
the thermal stability at all, at least at the order of a few decades. 
Robinson also pointed out the peculiarities of Nova V446 Her 1960. It 
is the only clear case of a classical nova that performed dwarf nova 
outbursts a few decades before the nova eruption and afterwards (Honeycutt 
et al. 1998). According to the 'modified hibernation scenario' V446 Her is 
different from other novae only by the time scales of the evolutionary 
processes.

An alternative view to the 'hibernation scenario' was presented by Mukai and 
Naylor (1993). They suggested that nova-likes and dwarf novae constitute
different classes of pre-nova systems. Therefore, there are two 
possibilities:\\
1. nova-likes $\rightarrow$ novae $\rightarrow$ nova-likes...\\
2. dwarf novae $\rightarrow$ novae $\rightarrow$ dwarf novae...\\
Nova-likes should have more frequent nova outbursts than dwarf novae because 
their mass ransfer rates are larger than those of dwarf novae, so the critical
mass for the thermonuclear runaway is achieved much faster.  
In the first option the thermal stability line is not crossed at all, while 
dwarf novae can cross the limit in the rare occasion of a nova outburst. It 
seems that the observations of old novae have not been able to judge between 
the 'modified hibernation scenario' and the different view. 

\section{A Method for Testing the Thermal Stability}

Retter and Leibowitz (1998) presented a very simple way to determine the 
thermal status of CVs, and particularly of old novae. We briefly summarize 
here this work. The equation for the thermal stability border-line is taken 
from Osaki (1996). It is assumed that the accretion disc is the dominant 
light source in the V band. The bolometric correction ($L_{V}/L_{bol}$) 
is estimated from models of stationary accretion discs (laDous 1989). 
A classical relation between the mass and radius of the white dwarf and 
typical white dwarf values are used for eliminating the radius of the 
white dwarf from the equations. The resulting two expressions are:\\
An equation for the critical V magnitude for crossing the thermal 
stability line:\\
\begin{equation}
(m_{V})_{crit}=2.16-4.25log(P_{orb})-3.33log(M_{wd}/M_{\odot})+5log(d)+A_{V}
\end{equation}
An equation for estimating the mass transfer rate:\\
\begin{equation}
\dot{M}/(10^{17}gr/sec)=(10^{\frac{m_{V}-A_{V}-0.69}{-2.5}})\frac{d^{2}}
{(M_{wd}/M_{\odot})^{4/3}}
\end{equation}
where $P_{orb}$ is the orbital period, $M_{wd}$ - the white dwarf mass, 
$M_{\odot}$ - the solar mass, d - the distance to the binary system and 
$A_{V}$ is the interstellar reddening in the V band. 

\section{The Early Presence of the Accretion Disc in Young Novae}

Even if the starting point of a pre-nova is in the lower, thermal instability 
zone as a dwarf nova, it is not clear that the nova eruption would carry 
the system into thermal stability, because the presence of an 
accretion disc is required. It was believed that the nova outburst destroys 
the disc, and that it takes only a few decades for the accretion disc to 
reform. However, observations of a few young novae, carried out at Wise 
Observatory, showed that in three cases the typical time scale for the 
re-establishment of the disc is weeks to very few years 
(Leibowitz et al. 1992; Retter, Leibowitz \& Ofek 1997; Retter, Leibowitz 
\& Kovo-Kariti 1998). It is therefore probable that the accretion disc even 
survives the nova event. Anyway, if a pre-nova is thermally unstable, the 
time scale for crossing the thermal stability limit can be very short.

\section{New Evolutionary Scenarios for Short Orbital Period CVs}

We focus now on short orbital period CVs. So far there are only two 
non-magnetic novae below the period gap - CP Pup 1942 and V1974 Cyg 1992. 
Both perform permanent superhumps in their light curves (Retter et al. 1997; 
Patterson \& Warner 1998). To these systems, we naturally add V603 
Aql 1918, the third permanent superhump nova (Patterson et al. 1997), 
whose binary period extends to the other side of the gap. (We note in 
passing that the division between tidally stable and unstable CVs is probably 
a little above the period gap).

Since all three systems are permanent superhumpers, there is no doubt that 
their accretion discs are thermally stable, and indeed a calculation, based 
on the equations presented above, confirms it. However, when we compare the 
pre-outburst luminosities with the post nova values, various types
of behaviour are discovered. V603 Aql seems to have returned exactly to its 
pre-outburst magnitude. The upper limit on the brightness of the 
progenitor of CP Pup (Warner 1995b) shows that it was fainter than its 
post outburst value, but prevents a precise decision concerning the thermal 
stability of the pre-nova. V1974 Cyg is the most interesting case among 
the three novae. It is relatively a young nova, which might still be decaying 
towards quiescence. Retter \& Leibowitz (1998) showed that the pre-nova was 
thermally unstable. It is thus the only clear case of a classical nova that 
changed its thermal stability state. Retter \& Leibowitz also anticipated 
two possible scenarios for the future status of the nova. It will either stay 
above a certain brightness level, keeping its permanent superhump state, or 
if decays below this limit, it will become thermally unstable again, turning 
into a regular SU UMa system. Such a transition has never been observed in 
any nova.

Retter \& Leibowitz also suggested that permanent superhump systems might be
ex-novae. Observational evidence for this idea comes for the possible 
identification of BK Lyn, a permanent superhump system (Skillman \& 
Patterson 1993), with a Chinese guest star, erupted in 101 (Hertzog 1986). 
We also note that two SW Sex candidates are old novae (Hoard 1998). Since 
permanent superhumps have been revealed in many SW Sex systems (Patterson 
1998), we regard this fact as another supportive evidence for this notion.

We further propose that evolutionary scenarios, similar to those offered for
the long orbital period CVs, are applicable to the short orbital period 
systems, too. The nova outburst is suggested as a mechanism for taking the 
accretion discs of SU UMa systems from the thermal instability zone into the 
stable part of the plane. After a few 
decades-millenia, the systems may decay back to their pre-outburst state. 
This scenario (SU UMa systems $\rightarrow$ permanent superhump systems 
$\rightarrow$ novae $\rightarrow$  permanent superhumpers $\rightarrow$ 
SU UMa systems...) resembles the 'modified hibernation scenario'. A 
different view, as mentioned in section 2, is that permanent superhump 
systems experience nova outbursts more often than SU UMa systems do. 
The two options are: \\
1. permanent superhump systems $\rightarrow$ novae $\rightarrow$ permanent 
superhump systems... \\
2. SU UMa systems $\rightarrow$ novae $\rightarrow$ permanent superhumpers 
$\rightarrow$ SU UMa systems... 

\section{Predicting the Presence of a New Type of Nova}

Classical novae are believed to occur on a binary system, that consists of a 
white dwarf and a red dwarf. In this section we speculate on the existence of 
a different kind of nova - a helium AM CVn like nova, which consists of two
degenerate helium white dwarfs. 

The orbital periods of AM CVn systems are very
short ($P_{orb}=17-50$ min.). Their spectra are rich in helium lines, but lack 
hydrogen lines. The standard model for these objects is that they consist 
of two degenerate helium white dwarfs, orbiting around each other in a very 
close orbit (Warner 1995b). During the last few years, superhumps have 
been found in five of the six known AM CVn members (Patterson 1998). 
Two systems are now believed to be in a permanent superhump state, while 
the other three are considered as helium dwarf novae. The disc instability 
model was successfully applied to these systems by Tsugawa \& Osaki (1997).

We continue the analogy between long and short orbital period CVs and apply it
to the ultra-short period AM CVn systems. We propose that the two AM CVn 
permanent superhumpers are thermally stable as regular permanent superhump 
systems and nova-likes, and that the three helium dwarf novae are the 
equivalents of SU UMa and U Gem systems, whose accretion discs are thermally 
unstable. We thus suggest that helium dwarf novae evolve through nova 
eruptions into the AM CVn permanent superhumpers. The proposed evolutionary 
link is cyclic. Alternatively, AM CVn permanent superhump systems experience 
more frequent nova events than helium dwarf novae. In both models, the 
presence of the new type of nova, an AM CVn-like nova, is unavoidable.

\section{Summary}

\begin{itemize}
\item Nova progenitors are usually thermally stable during the few decades
preceding and following their outbursts, but there are a few exceptions.
\item Observations show that the thermal stability line is indeed crossed 
during some nova outbursts and decays. It is not clear
whether this transition occurs in all non-magnetic novae or just a small 
part of the population.
\item In a few cases the thermal stability line is crossed very shortly after 
the nova eruption.
\item The time scales for crossing the line from top towards bottom is 
probably of the order of a few decades to a few millenia. It is also possible
that in some or most cases the novae stay thermally stable during all the 
time interval between successive nova outbursts.
\item Two evolutionary scenarios for short and ultra short orbital period 
CVs have been offered. They are similar to the scenarios for long orbital 
period CVs.
\item Based on the similarities between the different superhumping subclasses
of CVs, the presence of a new type of nova, an AM CVn-like nova, is suggested.
\end{itemize}



\section{References}

\vspace{1pc}


\re
1.\ laDous C.\ 1989, A\&A 211, 131
\re
2.\ Hertzog K.P.\ 1986, Observ. 106, No. 1071, 36
\re
3.\ Hoard D., 1998, Ph.D. thesis
\re
4.\ Honeycutt R.K., Robertson J.W., Turner G.W., Henden A.A.
1998, ApJ 495, 93313.\ Hoard D., 1998, Ph.D. thesis
\re
5.\ Leibowitz E.M., Mendelson H., Bruch A., Duerbeck H.W., 
Seitter W.C., Richter G.A.\ 1994, ApJ 421, 771
\re
6.\ Livio M.\ 1989, in Physics of Classical Novae, eds.\
Cassatella, Viotti (Springer-Verlag), p. 342
\re
7.\ Mukai K., Naylor T.\ 1995, in Cataclysmic Variables, eds.\ 
Bianchini, Della Valle, Orio (Kluwer Academic Publishers).
\re
8.\ Osaki Y.\ 1996, PASP 108, 39
\re
9.\ Patterson J.\ 1998, personal communication
\re
10.\ Patterson J., Kemp J., Saad J., Skillman D., Harvey D., Fried
R., Thorstensen J.R., Ashley R.\ 1997, PASP 109, 468
\re
11.\ Patterson J., Warner B.\ 1998, PASP 110, 1026
\re
12.\ Retter A., Leibowitz E.M., Ofek E.O.\ 1997, MNRAS 286, 745
\re
13.\ Retter A., Leibowitz E.M., Kovo-Kariti O.\ 1998, MNRAS 293, 145
\re
14.\ Retter A., Leibowitz E.M.\ 1998, MNRAS 296, L37
\re
15.\ Robinson E.L., 1975, AJ., 80, 515. 
\re
16.\ Shara M.M.\ 1989, in Physics of Classical Novae, eds.\
Cassatella, Viotti (Springer-Verlag)
\re
17.\ Skillman D.R., Patterson J.\ 1993, ApJ 417, 298
\re
18.\ Tsugawa M., Osaki Y.\ 1997, PASJ 49, 75
\re
19.\ Warner B.\ 1995a, Astrophysics \& Space Science 230, 83
\re
20.\ Warner B.\ 1995b, CV Stars (Cambridge University Press)




%

\chapter*{ Entry Form for the Proceedings }

\section{Title of the Paper}

{\Large\bf %
Novae crossing the thermal stability limit
}

\section{Author(s)}

I am the author of this paper.

\newcounter{author}
\begin{list}%
{Author No. \arabic{author}}{\usecounter{author}}

\item %
\begin{itemize}
\item Full Name:                Alon Retter
\item First Name:               Alon 
\item Middle Name:               
\item Surname:                  Retter  
\item Initialized Name:         A. Retter 
\item Affiliation:              Keele, U.K. 
\item E-Mail:                   ar@astro.keele.ac.uk 
\item Ship the Proceedings to:  Keele Uni., Staffs., U.K., ST5 5BG.
\end{itemize}


\end{list}

\end{document}